\documentclass[lengthcheck,twocolumns,pre,superscriptaddress,showpacs,floatfix,aps]{revtex4-2}

\usepackage{textcomp}
\usepackage{array,comment}
\usepackage{amsmath,amssymb,amsfonts,latexsym,bm}
\usepackage{graphicx,epsfig,caption,subcaption}
\usepackage{fancyvrb}
\usepackage{booktabs}
\usepackage{soul}
\usepackage{algorithm}
\usepackage{algpseudocode}

\usepackage{ragged2e}
\usepackage[svgnames]{xcolor}
\usepackage{xfrac}
\usepackage{mathrsfs}

\usepackage[colorlinks,linkcolor=MediumBlue,citecolor=MediumBlue,urlcolor = MediumBlue]{hyperref}
\usepackage{url}

\allowdisplaybreaks

\begin{document}

\title{Enhanced Gas Source Localization Using
Distributed IoT Sensors and Bayesian Inference}
\author{Leonardo Balocchi}\email{leonardo.balocchi@dottorandi.unipg.it}
\affiliation{Department of Engineering, University of Perugia, 06125 Perugia, Italy}
\author{Lorenzo Piro}
\affiliation{Department of Physics \& INFN, University of Rome ``Tor Vergata", Via della Ricerca Scientifica 1, 00133 Rome, Italy}
\author{Luca Biferale}
\affiliation{Department of Physics \& INFN, University of Rome ``Tor Vergata", Via della Ricerca Scientifica 1, 00133 Rome, Italy}
\author{Stefania Bonafoni}
\affiliation{Department of Engineering, University of Perugia, 06125 Perugia, Italy}
\author{Massimo Cencini} 
\affiliation{Istituto dei Sistemi Complessi, CNR, Via dei Taurini 19, 00185 Rome, Italy}
\affiliation{INFN ``Tor Vergata", Via della Ricerca Scientifica 1, 00133 Rome, Italy}
\author{Iacopo Nannipieri}
\affiliation{Sensichips s.r.l}
\author{Andrea Ria}
\affiliation{Dipartimento Ingegneria dell'Informazione, Università di Pisa, Via G. Caruso 16,56122 Pisa, Italy}
\author{Luca Roselli}
\affiliation{Department of Engineering, University of Perugia, 06125 Perugia, Italy}

\date{19-11-2024}

\begin{abstract}
Identifying a gas source in turbulent environments presents a significant challenge for critical applications such as environmental monitoring and emergency response. This issue is addressed through an approach that combines distributed IoT smart sensors with an algorithm based on Bayesian inference and Monte Carlo sampling techniques. Employing a probabilistic model of the environment, such an algorithm interprets the gas readings obtained from an array of static sensors to estimate the location of the source. The performance of our methodology is evaluated by its ability to estimate the source's location within a given time frame.
To test the robustness and practical applications of the methods under real-world conditions, we deployed an advanced distributed sensors network to gather water vapor data from a controlled source. The proposed methodology performs well when using both the synthetic data generated by the model of the environment and those measured in the real experiment, with the source localization error consistently lower than the distance between one sensor and the next in the array.
\end{abstract}

\maketitle

\section{Introduction to gas measurements challenge} 

Air pollution, both in urban and indoor environments, is a growing threat to public health, driven by rapid urbanization, increased vehicle emissions, and inadequate ventilation in modern buildings~\cite{gonzalez2021state,yan2021electrostatic}. In urban settings, emissions from vehicles and industrial activities release harmful pollutants such as Carbon Monoxide ($CO$), Carbon Dioxide ($CO_2$), and Nitrogen Dioxide ($NO_2$), which contribute to respiratory diseases, premature mortality, and environmental degradation~\cite{siregar2020air, chen2020intelligent}.
In indoor environments, especially during colder months, pollutant levels often rise due to energy-efficient insulation and limited ventilation in modern and refurbished buildings. This can lead to the accumulation of volatile and semi-volatile organic compounds, with concentrations often surpassing outdoor levels~\cite{dimitroulopoulou2012ventilation}. Poor Indoor Air Quality (IAQ) is linked to respiratory problems, allergies, and decreased cognitive performance, particularly in educational settings~\cite{marchetti2015campus, kumar2009indoor}. High levels of $CO_2$ concentrations, a marker of poor ventilation, are also associated with increased humidity and reduced productivity, highlighting the need for continuous IAQ monitoring.

\begin{figure}[t!]
    \includegraphics[width=1\columnwidth]{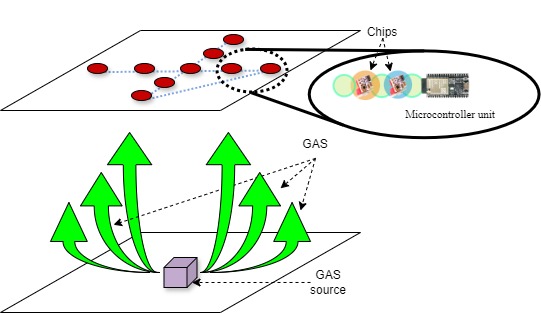}
    \caption{Gas measurement setup with an array of sensors (red circles) placed on the ceiling and connected to the Micro Controller Unit (MCU) via the communication protocol, indicated with the blue dash. The source is placed on the floor, but its exact position is unknown and must be inferred from the sensors' measurements.}
    \label{fig:scheme}
\end{figure}

In addition to air pollution, indoor gas safety is another critical concern, particularly with the use of natural gas in residential areas. Gas leaks pose significant hazards, including fire risks, necessitating timely detection and prevention. Traditional gas alarm systems, typically featuring a warning device and cut-off valve, are being upgraded with intelligent technology~\cite{hernandez2021recent}. The rise of "smart kitchens" and the integration of Internet of Things (IoT) technology have paved the way for more advanced gas monitoring systems. A key innovation is the use of Narrowband IoT (NB-IoT) gas meters, which provide real-time leak detection and wireless connectivity~\cite{jinfeng2021management}. These systems, when combined with comprehensive gas safety platforms, allow for monitoring, analysis, and remote management of gas-related risks. Indeed, identifying the source of harmful odors, such as those from gas leaks or chemical emissions, is essential for preventing environmental disasters~\cite{bayat2017environmental, burgues2020environmental}.
Recent advancements in air quality and gas safety monitoring thus represent significant strides toward creating healthier and safer indoor environments. By leveraging low-cost, portable sensor networks and smart technologies like NB-IoT, it is now possible to monitor pollutants such as $CO_2$, temperature, humidity, and gas levels in real time. Moreover, gas measurement sensors currently available are not optimized for the high-precision, distributed sensing that IoT applications demand~\cite{garrick2011comparison,ernst2001dynamic,dawei2011analysis,mishra2014fpga}.

On a more theoretical level, developing effective source localization algorithms and efficient data collection systems is a challenge of broad interest ranging from fundamental research with networks of static sensors for environmental monitoring applications~\cite{ferdinandi2019novel,tariq2021,esposito2022} to fluid dynamics~\cite{celani2014} and optimal navigation when considering mobile agents~\cite{piro2023_thesis,reddy2022_review}.
In this work, we propose an approach combining smart sensors~\cite{cerro2017preliminary,cerro2018metrological,bruschi2018novel,zampolli2024asic} and advanced algorithms based on Bayesian inference~\cite{johannesson2004,elfring2021,piro2024} to improve the localization of an indoor source of gas.
As a test case, we focus on identifying the position of a water vapor source placed on the floor of a room with a network of nine static sensors, as illustrated in Figure~\ref{fig:scheme}, that, after calibration, are used to collect data on the gas released by the source.
Such experimental data are then used to infer the source location using a Bayesian scheme implemented via a sequential Monte Carlo algorithm. 
To this end, one of the main challenges is modeling gas transport, which is essential when dealing with Bayesian methods. To mimic the poor control over the environmental conditions typical of realistic situations, we consider a rather simplified environment model that neglects any dynamical effect due to convective motions in the room, namely the so-called Gaussian plume model~\cite{sutton1932,hutchinson2017review}.
We show that, notwithstanding the simplifying assumptions, the gas source is localized with remarkable accuracy, as compared to the distance between the sensors, and rather quickly.\\

The paper is organized as follows. In Sec.~\ref{sec:experiment}, we illustrate the experimental setup and provide details about the sensors deployed and their calibration. 
Then, in Sec.~\ref{sec:numerics}, we introduce the model of the environment as well as the localization algorithm, namely the sequential Monte Carlo (SMC), used to interpret the sensors' detections and infer the gas source position.
We then present the results in Sec.~\ref{sec:results}, where, at first, we test the proposed algorithmic procedure using synthetic data generated from the same model of the environment used for the inference. Then, we study the performance of the same algorithm in localizing the source when using real data from the experimental setup. Finally, we summarize our findings and provide an outlook for future studies in Sec.~\ref{sec:discussion}.

\section{Experimental setup}\label{sec:experiment}

The actual data used for source localization are measured by a smart sensor platform and a new single-wire communication protocol, namely the Sensibus (see Figure~\ref{fig:scheme})~\cite{balocchi2024sensibus}. Inside the platform, a proprietary microanalytic mini-asic sensing platform equipped with on-chip sensing capabilities is embedded. It is called Sensiplus microChip (SPC).  Developed by Sensichips s.r.l. in
collaboration with the Department of Information Engineering of the University of Pisa~\cite{ria2021sensiplus}, SPC enables Electrochemical Impedance Spectroscopy (EIS) measurements on internal and external sensors~\cite{ria2023iot,gerevini2023end,bourelly2020preliminary,ria2023sensitag}. It is characterized by low power consumption (less than 1.4 mW), very small size (1.5 mm x 1.5 mm), and a potentially low cost (less than 10 USD in large-scale production). The SPC has no computational capacity~\cite{molinara2020end,cerro2017preliminary,bourelly2020chemicals} and is therefore complemented by a Micro Controller Unit (MCU) with (wired or wireless) communication capabilities.
Specifically, this advanced sensor platform, designed for IoT applications, is intended to perform real-time measurements and extend these readings across a broad spatial area. The connection between the SPC and MCU was established using the Sensibus protocol, a serial bus protocol that supports multiple SPCs connections, enabled by the SPC's 6-byte addressing capability.
Remarkably, there is an evident reduction in communication channel usage time when comparing a standard asynchronous single-wire data protocol, like 1-Wire, with Sensibus. A detailed description of the Sensibus protocol is provided in Ref.~\cite{balocchi2024sensibus}. Gas measurement is complex, requiring high precision and speed, especially to track gas movement and pinpoint its source. 
Nonetheless, the use of the Sensibus protocol, together with the SPC sensing features, is perfectly suited for this type of IoT objective. To this end, a specialized cable setup was created to connect the nine SPCs (see Figure~\ref{fig:scheme}), which are equipped with an aluminum oxide sensor that alters impedance based on exposure to gaseous substances, such as water vapor. 

\begin{figure}[t!]
    \centering
    \includegraphics[width=0.75\columnwidth]{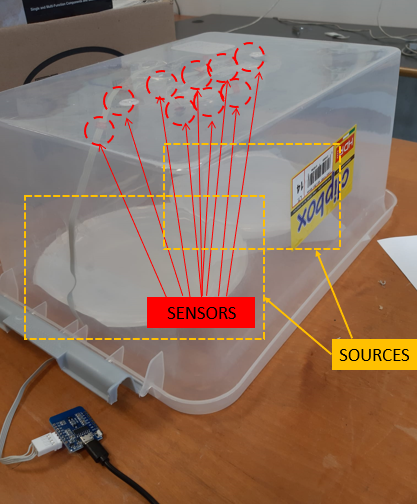}
    \caption{Calibration setup: the 9 sensor platforms (in the red circles) are enclosed in an environment with the same water vapor concentration, produced by two plastic plates containing water (in the yellow boxes).}
    \label{fig:calibsetup}
\end{figure}

The experiment was carried out in two steps. First, we exposed all the sensors to the same concentration of water vapor and looked at their response by switching on and off the water vapor source. In this way, we could properly calibrate the sensors, a necessary procedure since each of them has different amplitude responses when subjected to the same concentration of substance, due to variability in the manufacturing process of each chip.
Then, we moved on to the measurement phase, in which the nine sensors are placed in a room, left dangling from the ceiling, in the presence of a water vapor source on the floor. There, we could see how the sensors' response provides insightful information about the spatial distribution of the water vapor within the room that could then be used to locate the source by means of a suitable search algorithm.

\subsection{Sensors' calibration}

\begin{figure}
	\centering
	\subfloat[][]{
        \includegraphics[width=0.9\columnwidth]{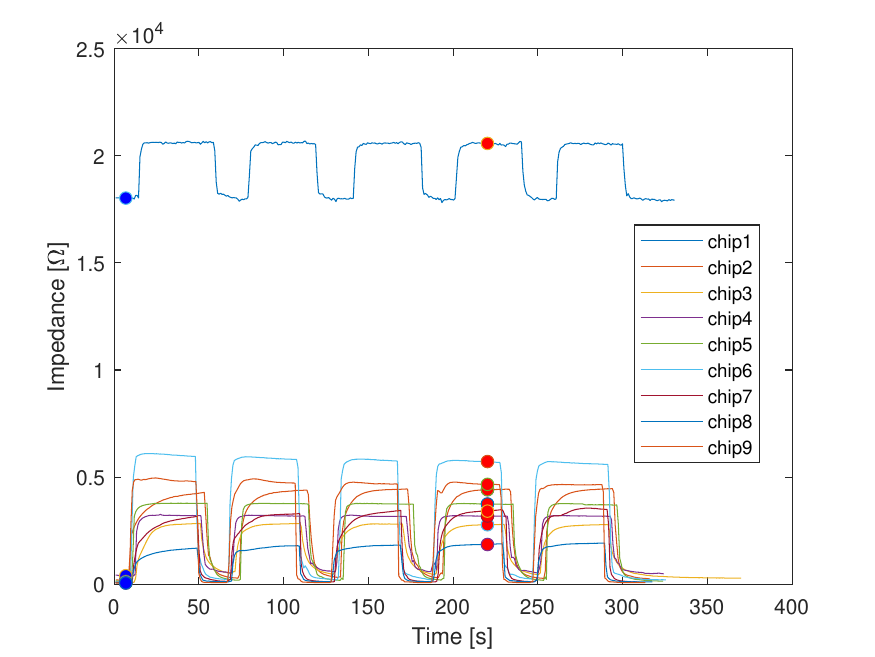}
        \label{fig:notcalib}
    }
 
    \hspace{2pt}
	\subfloat[][]{
        \includegraphics[width=0.9\columnwidth]{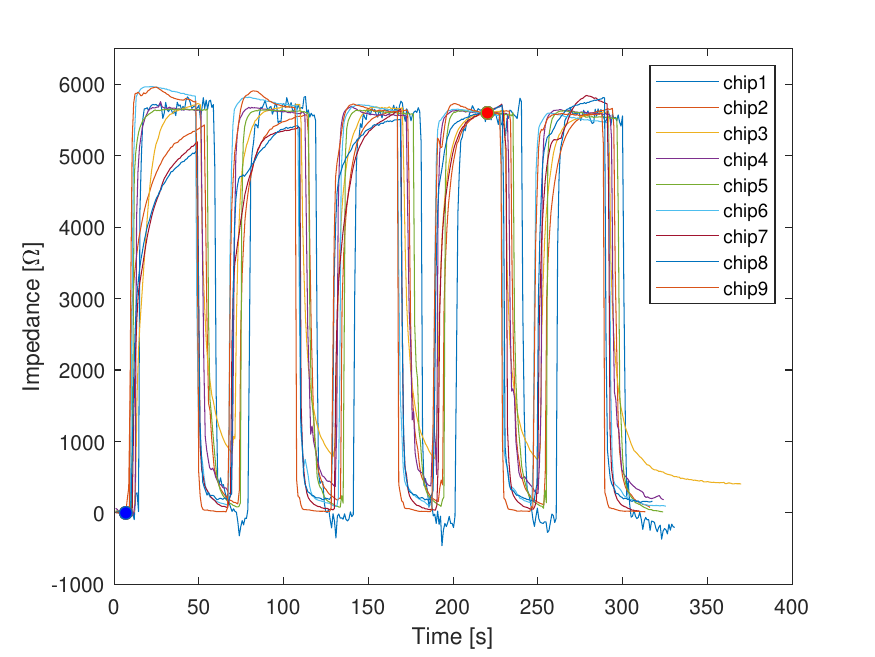}
        \label{fig:calib}
    }
    
    \caption{Calibration phase: measurement of aluminum oxide impedance module as a function of time for the nine SPCs. There are five periods of time in which the sensors are exposed to the water vapor (reaching saturation at the highest value), alternating with periods of time of non-exposure (low value). (a) Uncalibrated data; (b) same data calibrated using Eq.~\eqref{eq:calib} and the coefficients reported in Table~\ref{tab:protocol3} at a temperature of 26°C.}
 
\label{fig:calibAll}
\end{figure}

The fast real-time Sensibus communication protocol enables a simultaneous calibration phase for all the sensors as it allows for rapid collection of data from multiple sensors in minimal time. The calibration was carried out using a simple setup involving a plastic box, two water vapor sources to saturate the environment, and a commercial thermometer, as shown in Figure~\ref{fig:calibsetup}. Here, we adopt the following linear calibration formula:  
\begin{equation}
    cal_{\rm data} = \alpha*(raw_{\rm data}-\beta) \, ,
    \label{eq:calib}
\end{equation}
where $\alpha$ and $\beta$ are the coefficients used to align the baseline of the measurements, $raw_{\rm data}$ represents the SPC acquired data (expressed in $\Omega$), and $cal_{\rm data}$ stands for the calibrated data.  
Figure~\ref{fig:calibAll} shows the acquired raw data from each SPC in both the presence of water vapor concentration, indicated by the highest response of each sensor, and in the absence of it, corresponding to the lowest value of the impedance in the same plots.
The calibration experiment was conducted by alternating 40 seconds of exposure of the 9 SPCs to the same vapor concentration with 20 seconds of no exposure and repeating the procedure 5 times.
At this point, by means of Eq.~\eqref{eq:calib}, the lowest value of the impedance measured by each sensor was shifted to zero, while the highest value was rescaled to the average value among the ones of each sensor.
Figure \ref{fig:notcalib} shows how, without the calibration process, the 9 SPCs feature very different responses when exposed to the same input, due to differences in their manufacturing process.
In Figure~\ref{fig:calib}, the same signal after performing the calibration is reported, showing how the SPCs consistently have a similar response when exposed to the same concentration of water vapor.
Table \ref{tab:protocol3} shows the extracted $\alpha$ (adimensional) and $\beta$ (expressed in $\Omega$) coefficients used to normalize the measurements of the sensors, reported with an absolute error calculated over the 5 repetitions of each calibration experiment and for three different temperature values, as temperature affects the amount of vapor emitted from the source.

\begin{table*}[ht!]
\addtolength{\tabcolsep}{-1.5pt}
\centering
\caption{Extracted calibration coefficients for three different temperatures.}
\label{tab:protocol3}
\begin{tabular}{|c|c|c|c|c|c|c|}

\hline
\multicolumn{1}{|c|}{\textbf{}}&
\multicolumn{2}{|c|}{\textbf{Temperature 15°C}} & \multicolumn{2}{c|}{\textbf{Temperature 26°C}}  & \multicolumn{2}{c|}{\textbf{Temperature 40°C}}\\
\hline
\multicolumn{1}{|c|}{\textbf{Chip ID}}&
\multicolumn{1}{|c|}{\textbf{$\alpha$}} & \textbf{$\beta$} & \textbf{$\alpha$} &
\multicolumn{1}{c|}{\textbf{$\beta$}}& \textbf{$\alpha$} &
\multicolumn{1}{c|}{\textbf{$\beta$}}\\
\hline
1 & 2.3$\pm$0.1 & 18020$\pm$110 & 2.2$\pm$0.1 & 18020$\pm$110 & 2.2$\pm$0.7& 17660$\pm$115\\
2 & 2.50$\pm$0.09 & 67$\pm$3 & 1.4$\pm$0.1 & 322$\pm$5  & 2.1$\pm$0.1& 288$\pm$4\\
3 & 1.6$\pm$0.2 & 110$\pm$20 & 2.3$\pm$0.4 & 400$\pm$40& 1.9$\pm$0.2& 126$\pm$4\\
4 & 0.90$\pm$0.02 & 60$\pm$10 & 2.1$\pm$0.1 & 540$\pm$10& 2.1$\pm$0.9& 310$\pm$3\\
5 & 1.60$\pm$0.05 & 81$\pm$3 & 1.50$\pm$0.07 & 258$\pm$9& 1.20$\pm$0.03& 102$\pm$2\\
6 & 0.90$\pm$0.03 & 74$\pm$10 & 0.90$\pm$0.04 & 210$\pm$20& 0.90$\pm$0.02& 42$\pm$1\\
7 & 4.0$\pm$0.2 & 58$\pm$6 & 1.7$\pm$0.2 & 122$\pm$8& 2.1$\pm$0.1& 150$\pm$5\\
8 & 3.5$\pm$0.2 & 51$\pm$9 & 3.3$\pm$0.2 & 103$\pm$6& 0.90$\pm$0.03 & 111$\pm$5\\
9 & 1.30$\pm$0.04 & 140$\pm$7 & 1.10$\pm$0.03 & 75$\pm$4& 1.20$\pm$0.02& 237$\pm$4\\

\hline
\end{tabular}
\end{table*}

\begin{figure}
    \centering
    \includegraphics[width=0.9\columnwidth]{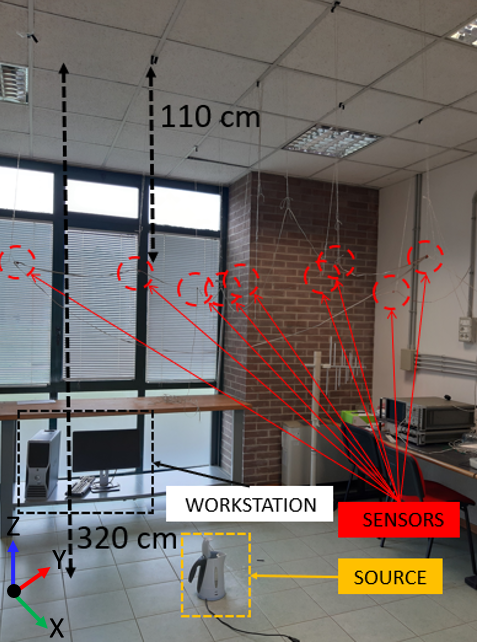}
    \caption{Setup of the gas measurement experiment, with the nine sensors placed above the source represented by a kettle with water. The workstation is connected to the MCU for storage and processing of the extracted data.}
    \label{ceilingSetup}
\end{figure}

\subsection{Distributed measurements} 

The combination of the Sensibus protocol and SPCs enables efficient analysis of various characteristics and distribution patterns of the gas being studied. During this second phase, the nine sensor platforms, all connected with a single cable, were arranged in a cross formation above a water vapor source, represented by a kettle full of water placed on the floor. As shown in Figure~\ref{ceilingSetup}, the SPCs were not strictly attached to the ceiling (ceil height is 320 cm) but at a fixed distance from it (110 cm). In fact, if we had placed them directly to the ceiling, the gas level measured by the sensors would saturate because of the water vapor accumulation over time in a layer close to the ceiling, and thus bias their response in the long term.

Different phases characterize the experiment. First, there is a period of time (about 5 minutes) when we have to wait for the water to reach the boiling temperature. Then, once the water in the kettle reached the boiling temperature, we started the measurement phase for a time equal to 20 minutes, during which the kettle continuously emitted water vapor. Finally, the kettle was turned off, and its lid closed to prevent the residual vapor stream from escaping. These measurements are then saved and processed in a workstation consisting of a computer connected to the MCU. 

\begin{figure}
   \centering
   \subfloat[][]{
        \includegraphics[width=0.9\columnwidth]{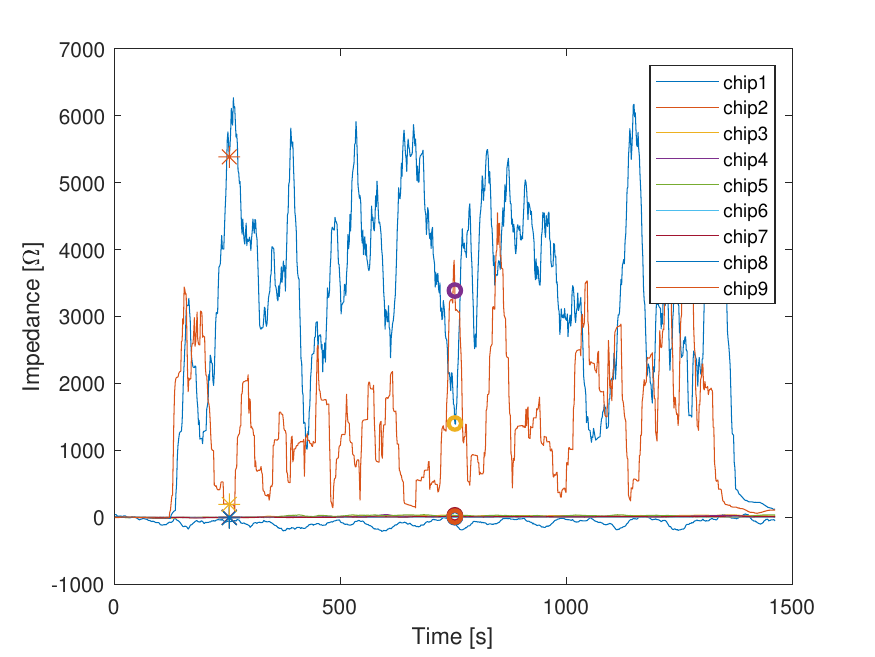}
        \label{fig:time_series}
    }
    \hspace{2pt}
	\subfloat[][]{
        \includegraphics[width=0.9\columnwidth]{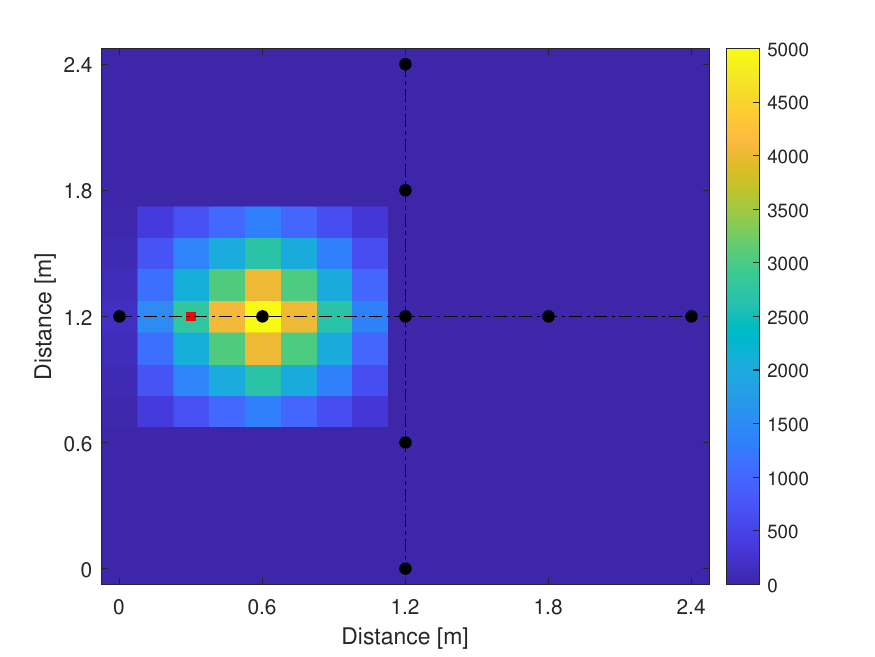}
        \label{fig:snapshotA}
    }
    \hspace{2pt}
	\subfloat[][]{
        \includegraphics[width=0.9\columnwidth]{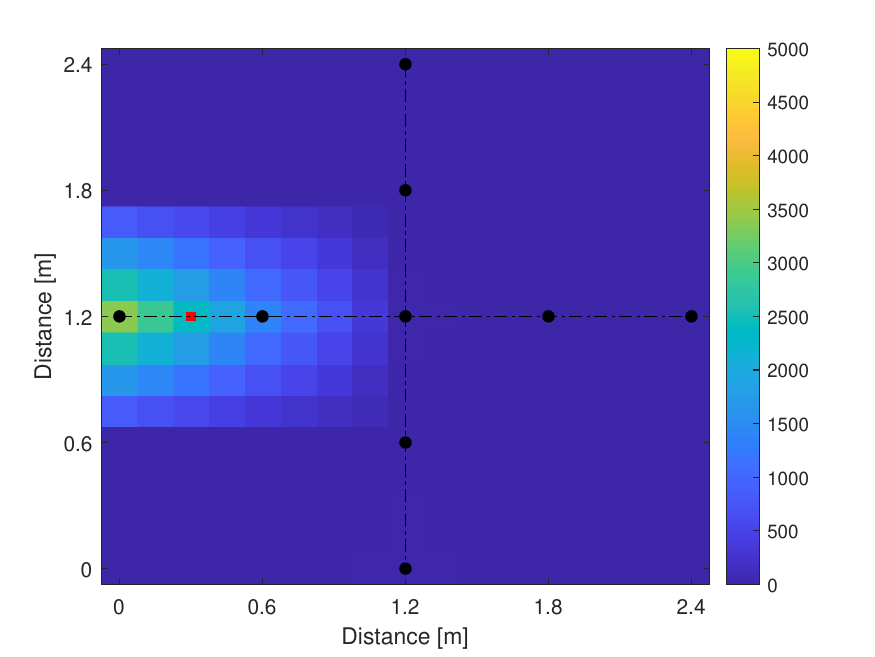}
        \label{fig:snapshotB}
    }
    \caption{
    Measurement of aluminum oxide impedance module over time for the 9 SPCs. (a) Red stars and red circles indicate the samples at time instant 250.5 s and 750 s.
    Maps of the water vapor spatial distribution at the same time instant (b) 250.5 s and (c) 700 s, obtained using linear interpolation. The black circles represent the sensors' locations, 60 cm apart, while the red square indicates the location of the source.}
    \label{fig:spat_meas}
\end{figure}

The rapid data collection from all the SPCs, as previously described, allowed for precise measurements of both spatial and temporal variations in the water vapor concentration at the sensors' positions.
The experiment in Figure~\ref{fig:spat_meas} depicts a typical acquisition, with the gas source shifted to the left with respect to the center of the sensor cross and indicated by the red square in Figures~\ref{fig:spat_meas}(b) and (c). The room temperature in the experiment shown is 26 °C. More specifically, Figure~\ref{fig:spat_meas}(a) shows the time-evolution in the sensors’ responses, measured as impedance values, while Figs.~\ref{fig:spat_meas}(b) and (c) present two temporal snapshots of the water vapor concentration measured by the sensors, as indicated by the corresponding symbols in the time signal in Figure~\ref{fig:spat_meas}(a).
Please note that these density maps have been obtained by a simple linear interpolation of the values of the concentration measurements made by the sensors. Therefore, these spatial distributions are represented here just for the sake of illustration and do not represent the true water vapor concentration field. In fact, the actual model of the environment deployed in our analysis is somewhat more complex and shall be presented in the next section.

\section{Numerical methods}\label{sec:numerics}

\subsection{Dispersion model}

To localize the vapor source position relying on sensors' measurements, we need to introduce a model that accounts for the mean-flow and mean-diffusion properties of the environment. 
We model the turbulent transport of a buoyant gas plume released at a rate $Q$ by a point source using a Gaussian plume model~\cite{hutchinson2017review,sutton1932}. Since we assume to know the distance of the source in the buoyancy direction $\hat{\bm{z}}$, the source localization problem reduces to a 2--D search in the plane where the sensors are placed, which is perpendicular to $\hat{\bm{z}}$ (see Figure~\ref{fig:scheme}). In the stationary regime, the analytical solution for the gas concentration in such a plane at known distance $z_{\rm s}$ from a source located in $\bm{r}_{\rm s}=(x_{\rm s}, y_{\rm s})$ is~\cite{sutton1932}
\begin{equation}
    c(\bm{r}-\bm{r}_{\rm s}) = \frac{Q}{\pi u \lambda^2} e^{-\frac{\lVert\bm{r}-\bm{r}_{\rm s}\rVert}{\lambda^2}} \, ,
    \label{eq:conc}
\end{equation}
where $Q$ is the emission rate, $u$ is the magnitude of the mean velocity of the plume in the vertical $\hat{z}$ direction, and $\lambda$ is the effective length scale of diffusion encompassing both molecular and turbulent diffusivity. 

Then, since water vapor is naturally present in the air, to limit the effect of the background noise in the signal, we discretize the gas concentration measured by each sensor into a finite number of levels, such that the probability of making a detection within a time interval $\Delta t$ takes the form~\cite{ristic2016}:
\begin{equation}
    p(h_i|d_i) = \frac{[\mu(d_i)]^{h_i}e^{-\mu(d_i)}}{h_i!} \, ,
    \label{eq:poisson}
\end{equation}
where $h_i=\{0,\dots,h_{\rm max}\}$ and $d_i \equiv \lVert\bm{r}_i-\bm{r}_{\rm s}\rVert$ correspond to the measurement made by the $i-$th sensor and its distance from the source, respectively, while $\mu$ represents the mean hit rate which can be straightforwardly obtained from the concentration via the Smoluchowski relation~\cite{smoluchowski1918versuch}:
\begin{equation}
    \mu(d_i)= \frac{\tilde{Q}}{\lambda^2}a\Delta t e^{-\frac{d_i^2}{\lambda^2}} \, ,
    \label{eq:mu}
\end{equation}
where $a$ is the sensor's detection radius, and $\tilde{Q}$ is an effective emission rate. 
Despite this oversimplification, this model can still capture the mean field characteristics of the buoyant plume of water vapor. One of its advantages is that considering Poissonian detections would allow us to generalize our approach to work also in turbulent environments where the gas signal is made sparse and intermittent due to turbulent transport~\cite{celani2014,vergassola2007}.
More importantly, another key benefit of this model is that it relies on only two environmental parameters, namely the effective source emission rate $\tilde{Q}$, and the characteristic length scale of the gas dispersion $\lambda$, which simplifies the parameters space exploration when trying to infer the source location.

\subsection{Source localization via Bayesian inference}

Each sensor measurement provides information about the source position $\bm{r}_{\rm s}$, which we process using Bayesian inference~\cite{bayes}. The combined measurements from all sensors update a probability map —-referred to as \emph{belief} in Bayesian jargon—- of the source's location inside the arena $\Omega$ (i.e. the probability domain), defined as $B(\bm{r})\equiv \mathrm{Prob}(\bm{r}_{\rm s}=\bm{r})$. Initially, the belief is set to a uniform distribution as we assume no prior knowledge about the source position.

Since the measurements from the $N_{\rm s}$ sensors at any time $t$ are independent, the conditional probability for a set of observations $\bm{h}^{(t)}$ at a potential source position $\bm{r}$ is given by:
\begin{equation}
    \mathcal{L}(\bm{h}^{(t)}|\bm{r}) = \prod\limits_{i=1}^{N_{\rm s}}p(h_i^{(t)}|d_i) \, ,
    \label{eq:likelihood}
\end{equation}
which depends on the environmental model specified in Eqs.~\eqref{eq:poisson}-\eqref{eq:mu} and is known as the \emph{likelihood} in Bayesian terms.
At each time step, after all sensors have measured, the belief is updated using Bayes' rule~\cite{box2011}:
\begin{equation*}
    B^{(t)}(\bm{r}) = \frac{\mathcal{L}(\bm{h}^{(t)}|\bm{r}) B^{(t-1)}(\bm{r})}{\int_\Omega {\rm d}\bm{r'} \, \mathcal{L}(\bm{h}^{(t)}|\bm{r'}) B^{(t-1)}(\bm{r'})} \, .
\end{equation*}
However, on the one hand, this path is not pursuable in practice since the computation of the integral at the denominator becomes quickly impractical from a computational point of view as the domain size grows. Moreover, we have to estimate online also the values of the unknown model parameters $\tilde{Q}$ and $\lambda$ that fit best with the sensors' measurements, which, together with the source position, makes the belief effectively four-dimensional.

To this end, we shall employ a state-of-the-art Sequential Monte Carlo (SMC) algorithm, enhanced with a Markov chain Monte Carlo (MCMC) perturbation step, to estimate the location of a gas source from sensor detection history~\cite{johannesson2004,elfring2021}. This approach allows simultaneous inference of the source position $\bm{r}_{\rm s}$ and the unknown parameters $\bm{P}=\{\tilde{Q},\lambda\}$ of the model of the environment while avoiding the numerical --and potentially very time-consuming-- computation of the integrals mentioned above.
For all the details concerning the SMC algorithmic implementation, please refer to the Appendix.

\section{Results}\label{sec:results}

Before implementing the SMC algorithm on the real data obtained in the experimental setup described in Sec~\ref{sec:experiment}, we run some preliminary tests.
In fact, importantly, once the algorithm’s hyperparameters are properly fine-tuned, SMC shall converge to the correct source location, assuming the environmental model is correctly specified.
However, this is no longer guaranteed when we do not have a perfect description of the environment, which is always the case in any application, and inaccuracies may arise when this is used to localize a gas source in realistic scenarios.

This is why we have devised a numerical experiment that emulates the real one, except that the sensors' measurements are generated from the simplified model of the environment described in the previous section (see Eqs.\eqref{eq:conc}-\eqref{eq:mu}) rather than from true gas concentration readings.

\subsection{Source localization in the numerical experiment}

\begin{figure*}[t!]
    \centering
    \includegraphics[width=\textwidth]{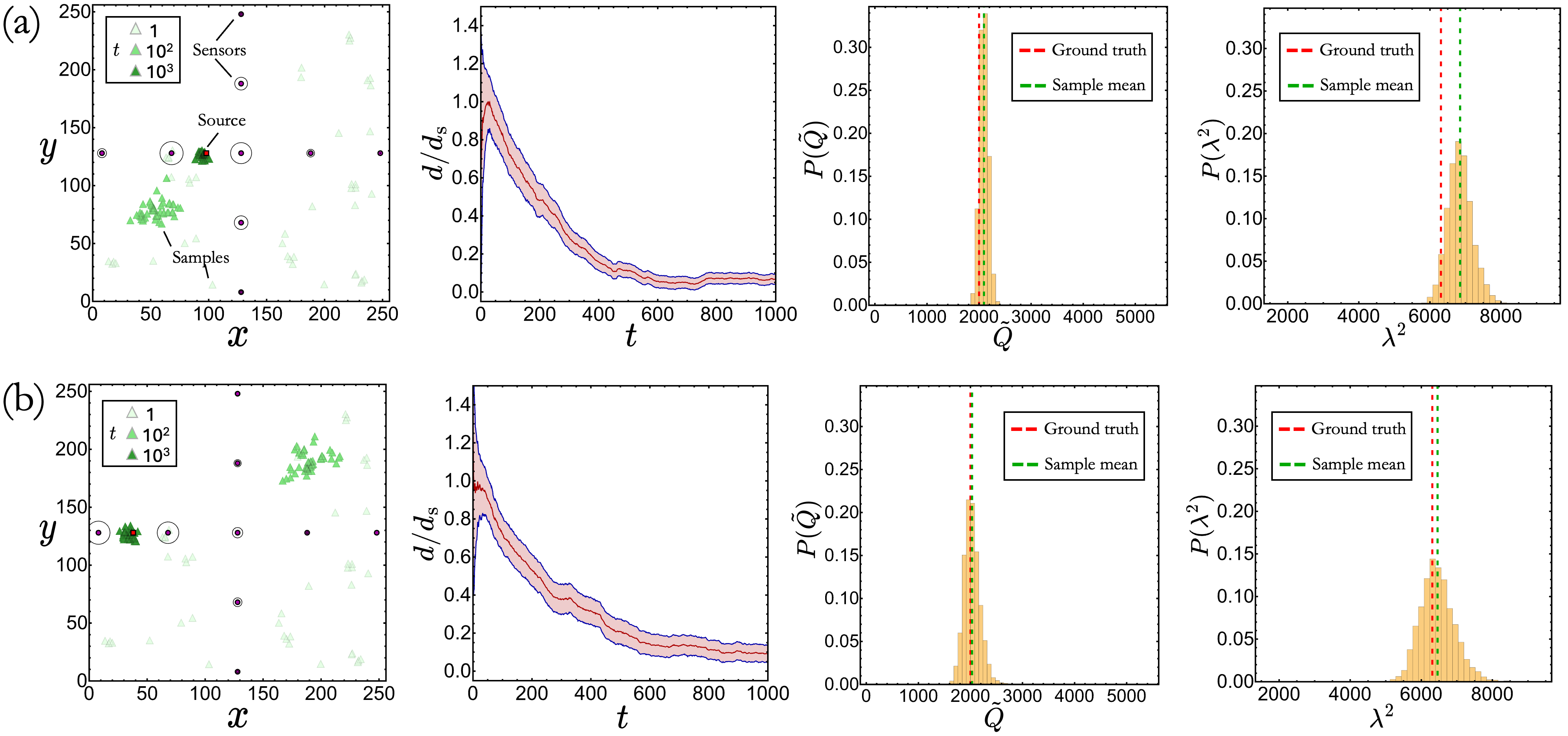}
    \caption{Results obtained in the numerical experiment. Starting from the left, the figures represent two spatial plots related to the two different source locations indicated by the red square, whose $(x,y)$ coordinates are for line (a) (98, 128) and for line (b) (38, 128). The 9 purple circles spaced evenly, represent the 9 sensors (Sensiplus) used for the measurements. By contrast, the triangles of different shades of green, going from the lightest to the darkest, represent the sequential Monte Carlo (SMC) samples at increasing time instants, from 1s up to 1000s. The black circles around each sensor are proportional to the total number of detections made by that sensor during the experiment.
    The plots in the second column represent the distance $d$ between the center of mass of the SMC samples and the true position of the gas source, normalized with respect to the value of the constant distance $d_{\rm s}$ between the sensors.  The data shown here for each source are obtained from an average over $10^3$ independent realizations, with the solid red line being the mean value and the shaded area between the blue curves a confidence interval given by the standard deviation. Lastly, the histograms in the right-hand graphs show the distributions of the model's free parameters $\lambda^2$ and $\tilde{Q}$, with red dashing being the ground and the green dashing the values estimated by SMC.}
\label{fig:notreal_data_mcm}
\end{figure*}

To conduct the numerical test, we have defined a search space where we aim to locate our source, corresponding to a square arena of size 256 x 256 $\rm cm^2$. We have then positioned nine sensors in a way to replicate the arrangement used in the actual laboratory setup, as indicated by the purple circles in the leftmost panels of Figure~\ref{fig:notreal_data_mcm}. Consequently, we have performed two distinct experiments by placing the gas source in two different locations, as indicated by the red squares in the same panels, with one source located between the central sensor and the one to its left (Figure~\ref{fig:notreal_data_mcm}(a)), while the second source lies between the second sensor on the left and the farthest sensor on the left (Figure~\ref{fig:notreal_data_mcm}(b)).
Note that these source locations are the same ones used in the real experiments, which we will discuss in the next subsection.

We have then set the environmental parameters (our ground truth in the numerical experiment) to $\lambda=80 {\rm cm}$ and $\tilde{Q}=2\cdot10^3 {\rm mol\cdot cm/s}$, and run the SMC algorithm starting from a uniform prior distribution for both the source location $\bm{r}_{\rm s}$ and such two free parameters of the model, $\lambda$ and $\tilde{Q}$. Details on the SMC hyperparameters used in this setup, as well as in the real experiment, can be found in Table~\ref{tab:hyper_SMC} in the Appendix. 

Each numerical simulation has a fixed duration of $10^3$ seconds, with each sensor collecting one measurement per second. 
As shown by the three temporal snapshots of the SMC samples in the leftmost panel in Figure~\ref{fig:notreal_data_mcm} (green triangles), the SMC algorithm progressively converges towards the true source location over time. 

To make this statement more quantitative and meaningful, we have then repeated the same two experiments for $10^3$ times each. As a result, we can compute the time evolution of the distance $d$ between the estimated source position and the true one as a function of time, averaged over all the $10^3$ independent realizations of each experiment, as shown in the second column of panels in Figure~\ref{fig:notreal_data_mcm}.
Such a distance (in units of the one between sensors, $d_{\rm s}$) consistently goes to zero over the total time of the simulations for both source locations. This highlights the accuracy and reliability of the SMC algorithm deployed for our source localization task. In particular, considering the final $d/d_{\rm s}$ ratio, the source localization error is close to $10\%$.

Furthermore, we could observe that the algorithm also correctly estimates the values of the model parameters that define the environment, namely $\lambda$ and $\tilde{Q}$. Indeed, since we know the ground truth, we can assess the accuracy of these estimates. 
The parameter distributions obtained from all our numerical simulations are shown in the third and fourth columns in Figure~\ref{fig:notreal_data_mcm}. Remarkably, the ground truth, indicated by a red dashed line, consistently falls within the range of values provided by the algorithm and is always very close to the sample mean, shown as a green dashed line. 

These results demonstrate that the algorithm can successfully estimate both the source location and the model’s parameters even though starting the search without any prior knowledge of the environment.

\begin{figure*}[t!]
	\centering
\subfloat[][]{
        \includegraphics[width=0.8\columnwidth]{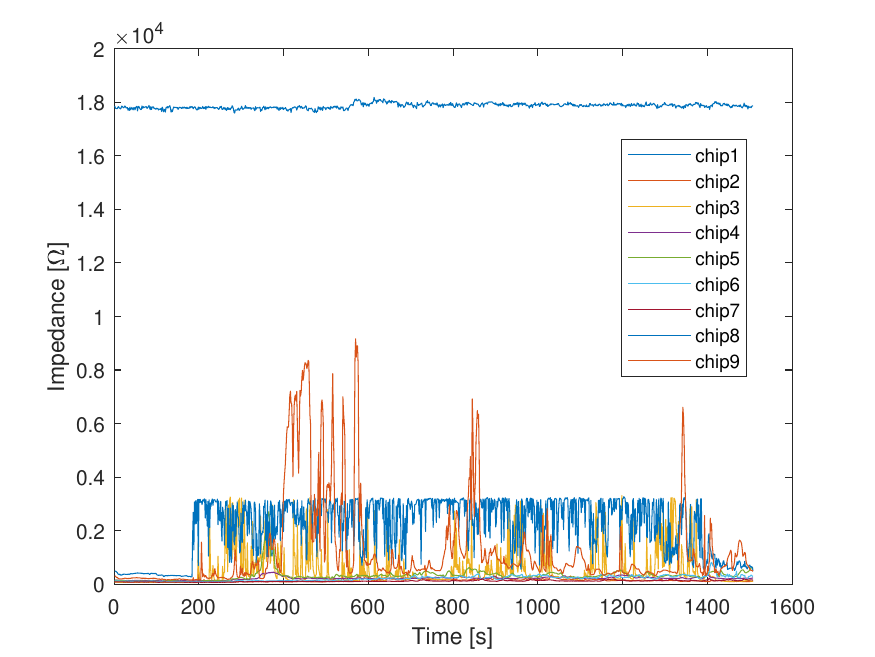}
        \label{fig:snapshotC}
    }\quad\subfloat[][]{
        \includegraphics[width=0.8\columnwidth]{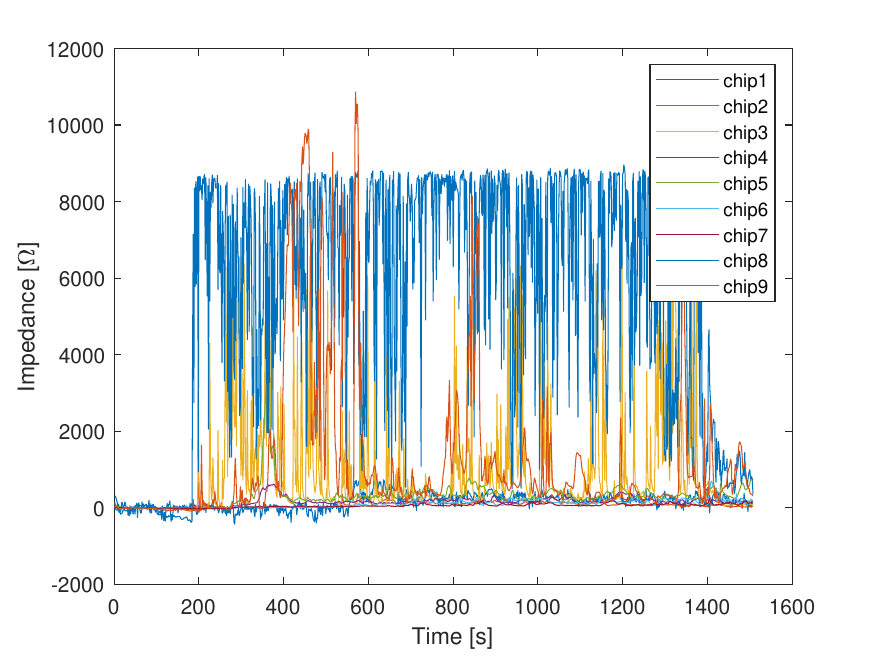}
        \label{snapshotD}
    }
    
    \hspace{1pt}
    
    \subfloat[][]{
        \includegraphics[width=0.8\columnwidth]{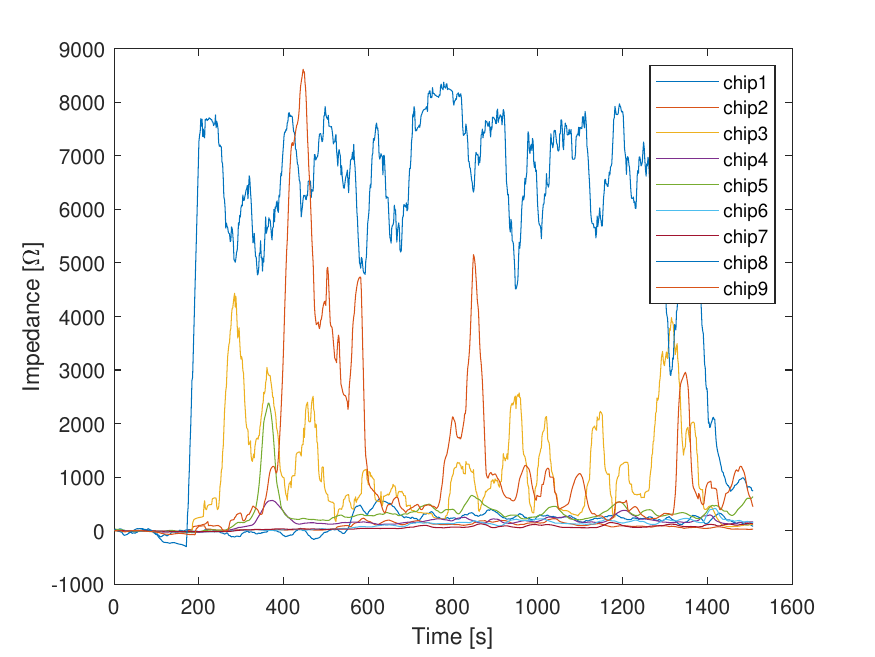}
        \label{snapshotE}
    }\quad\subfloat[][]{
        \includegraphics[width=0.8\columnwidth]{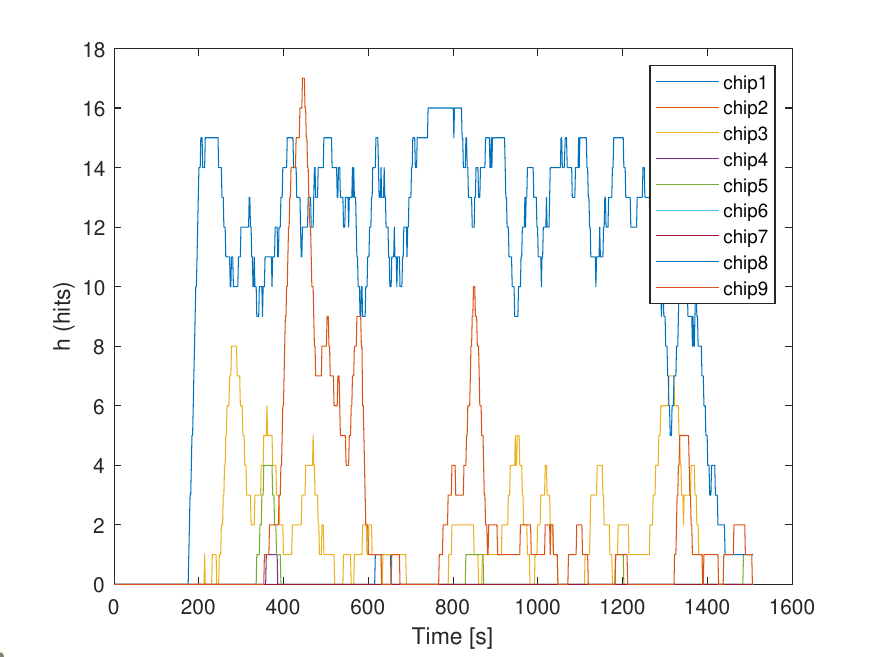}
        \label{snapshotF}
    }
    
    \caption{Data corresponding to the experiment with the source placed between the middle and left sensors, i.e. $\bm{r}_{\rm s} = (98,128)\rm cm$. (a) Raw data directly obtained from the sensors' measurements. (b) Same data as in (a) after calibration. (c) Moving average of the calibrated data shown in (b) using a time window of 20 seconds. (d) Discretized signal (number of hits $h$) obtained using a threshold of 500 $\Omega$ that can be used in the SMC algorithm to infer the source position.}
    \label{fig:measurements}
\end{figure*}

\subsection{Source localization in the real experiment}

\begin{figure}[ht!]
    \centering
    \includegraphics[width=\columnwidth]{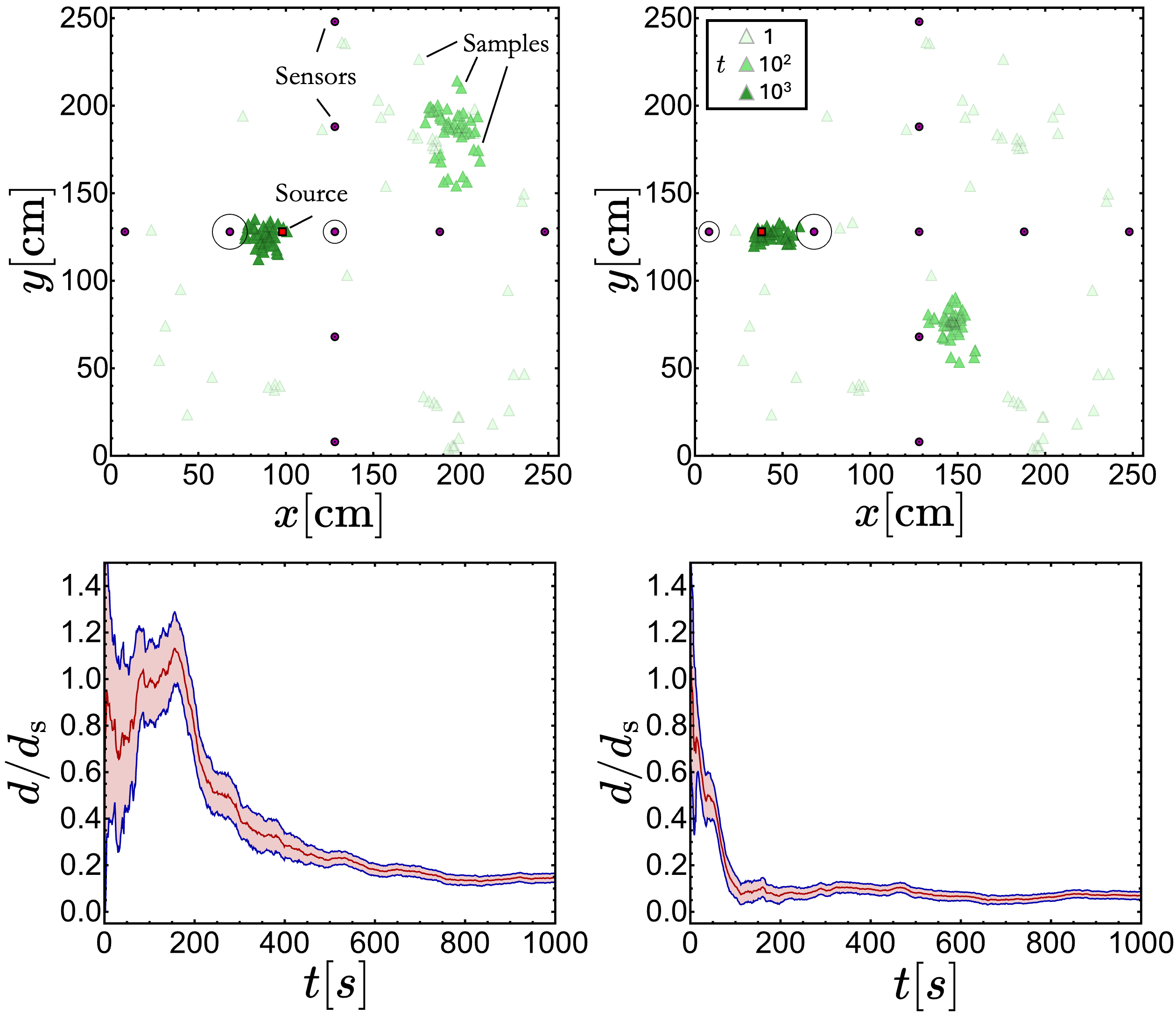}
    \caption{Results obtained in the real experiment. Starting from the top, there are two spatial plots related to the two different source locations indicated by the red square, whose $(x,y)$ coordinates are (98, 128) for the left column and (38, 128) for the right column. The nine evenly-spaced purple circles represent the sensors (Sensiplus) used for the measurements. The triangles of different shades of green, going from the lightest to the darkest, represent the sequential Monte Carlo (SMC) samples at increasing time instants, from 1s up to 1000s.
    The black circle around each sensor is proportional to the total number of detections made by that sensor during the experiment.
    The plots at the bottom show the distance $d$ between the center of mass of the SMC samples and the true position of the gas source, normalized with respect to the value of the constant distance $d_{\rm s}$ between the sensors. The data shown here for each source are obtained from an average over 10 experiments, with the solid red line being the mean value and the shaded area between the blue curves a confidence interval given by the standard deviation.}
    \label{fig:real_data_mcm}
\end{figure}

The algorithm reported in Sec.~\ref{sec:numerics} has been applied to the experimental measurements acquired with the setup described in Sec.~\ref{sec:experiment} for source localization.
To verify the repeatability of the experiment and to provide a sufficiently large data set over which we could average our results, the same experiment was repeated 10 times for each of the two source locations considered -- namely, the same ones used in the numerical experiment described in Sec.~\ref{sec:results}A. 
However, before implementing the SMC algorithm to infer the source location, we had to properly preprocess the raw data obtained from the sensors' measurements (see Figure~\ref{fig:measurements}a). First, after the application of the calibration formula (Figure~\ref{fig:measurements}b), we computed a moving average so as to smoothen the data and make the signal less noisy (see Figure~\ref{fig:measurements}c). Then, we thresholded the signal to transform the continuous sensors' measurements into discrete detections -- or \emph{hits} $h$, as shown in Figure~\ref{fig:measurements}d-- that could be finally used by the SMC algorithm, as described in Sec.~\ref{sec:numerics}.

In this realistic setting, the model deployed to interpret the sensors' measurement only approximates the physical laws governing the environment and, given its oversimplification, there does not exist a combination of the parameters $\lambda$ and $\tilde{Q}$ that can exactly describe the water vapor dispersion.
Consequently, we are not interested in inferring
the values of such parameters. Instead, we aim to evaluate whether the algorithm could still accurately identify the source's location, despite using an inaccurate model.
Remarkably, the algorithm deployed managed to locate the source with high accuracy in both the configurations considered. Indeed, as reported in Figure~\ref{fig:real_data_mcm}, we can see that the Monte Carlo samples still converge in proximity to the gas source (top panels). More quantitatively, we found that the average distance between the estimated and actual source locations goes down to $10-20\%$ of the distance between sensors (equal to 60 cm) by the end of the experiment. Furthermore, although the experiments lasted for about 20 minutes ($1600$ seconds), the search was stopped much earlier (after $1000$ seconds) since the algorithm could converge to the correct source location typically within such a shorter time frame.

\section{Conclusions}\label{sec:discussion}

We successfully developed a comprehensive IoT-based methodology for gaseous source localization. This approach integrates both hardware, through the careful selection of sensor types and communication protocols, and software, by implementing a search algorithm first validated using synthetic data and finally applied to real experiments. To the best of our knowledge, this is one of the few results based on real data processed with MC methods for the search of gaseous source~\cite{wang2017locating,8629026}.
This work shows quite promising results in localizing the gas source, with an accuracy of $10-20\%$ of the distance between the sensors. The results have been consistently tested over repeated experiments.

Concerning possible future extensions and improvements of the developed methodology, two possible paths are certainly worth investigating. First, the hardware can be improved by wirelessly replicating the Sensibus protocol, thus making the sensor network more versatile and easy to set up, even on mobile platforms such as drones. Secondly, on the numerical side, it is possible to deploy more sophisticated search algorithms that are suited to exploit multiple models as, e.g., the recently introduced Weighted Bayesian update~\cite{piro2024}, especially in more complex environments where the environmental model is even less under control.

\section*{Acknowledgments}

This work was supported in part by the Italian Ministry of University and Research (MUR), in the frame of the “PON 2022 Ricerca e Innovazione” action, in part by the European Research Council (ERC) under the European Union’s Horizon 2020 research and innovation program (Grant Agreement No. 882340) and in part by the EU Horizon Europe RHE-MEDiation with (GA 101113045).
L. Biferale and M. Cencini acknowledge financial support under the National Recovery and Resilience Plan (NRRP), Mission 4, Component 2, Investment 1.1, Call for tender No. 104 published on 2.2.2022 by the Italian Ministry of University and Research (MUR), funded by the European Union – NextGenerationEU– Project Title Equations informed and data-driven approaches for collective optimal search in complex flows (CO-SEARCH), Contract 202249Z89M. – CUP B53D23003920006 and E53D23001610006.

This work was supported in part by the Italian Ministry of University and Research (MUR), in the frame of the “PON 2022 Ricerca e Innovazione” action, in part by the European Research Council (ERC) under the European Union’s Horizon 2020 research and innovation program (Grant Agreement No. 882340) and in part by the EU Horizon Europe RHE-MEDiation with (GA 101113045).
L. Biferale and M. Cencini acknowledge financial support under the National Recovery and Resilience Plan (NRRP), Mission 4, Component 2, Investment 1.1, Call for tender No. 104 published on 2.2.2022 by the Italian Ministry of University and Research (MUR), funded by the European Union – NextGenerationEU– Project Title Equations informed and data-driven approaches for collective optimal search in complex flows (CO-SEARCH), Contract 202249Z89M. – CUP B53D23003920006 and E53D23001610006.

\section*{Abbreviations}{
The following abbreviations are used in this manuscript:}

\noindent 
\begin{tabular}{@{}ll}

IAQ & Indoor Air Quality\\
IoT & Internet of Things\\
NB-IoT & Narrowband IoT\\
SMC & Sequential Monte Carlo\\
IAQ & Indoor Air Quality\\
SPC & Sensiplus microChip\\
MCU & Micro Controller Unit\\
MCMC & Markov chain
Monte Carlo\\

\end{tabular}

\vspace{11pt}

\appendix
\setcounter{figure}{0}
\renewcommand\thefigure{A\arabic{figure}}
\setcounter{equation}{0}
\renewcommand\theequation{A\arabic{equation}}
\setcounter{table}{0}
\renewcommand\thetable{A\arabic{table}}

\section{Details on the numerical simulations}

Let us briefly outline what the SMC algorithm entails. 
At each time step, once all sensors have performed the measurement, we draw $N$ samples $\bm{\theta}_i \equiv \{\bm{r}_{{\rm s},i},\bm{P}_i\}$ from the current belief in the parameter space. Each sample is assigned a weight $w_i$ based on the likelihood of the latest measurement. We normalize the weights and calculate the effective sample size $N_{\rm eff}$ to address the degeneracy problem~\cite{elfring2021}. If \(N_{\rm eff}\) falls below a threshold (typically $N/2$), we perform a resampling step where each sample is replaced with a probability equal to its current weight, and all new weights are then updated to $1/N$.

Subsequently, we apply the Metropolis-Hastings MCMC perturbation, generating a Markov chain of length $K$ for each sample. More specifically, new inferences $\hat{\bm{\theta}}^{(t)}$ are drawn using a Gaussian proposal distribution $q(\hat{\bm{\theta}}^{(t)}|\tilde{\bm{\theta}}_{i,j-1}^{(t)}) = \mathcal{N}(\tilde{\bm{\theta}}_{i,j-1}^{(t)},\sigma^2)$. Then, the acceptance ratio, incorporating the likelihood history and proposal ratios, determines whether to accept the new sample~\cite{johannesson2004}.
Following such perturbation, the updated belief is approximated as $\tilde{B}^{(t)}(\bm{\theta}) = \frac{1}{N} \sum_{i=1}^N \delta(\bm{\theta}-\bm{\theta}^{(t)}_i)$. For more details, please refer to the pseudo-code reported in Algorithm~\ref{alg:SMC}.

\begin{table}[!ht]
\caption{Values of the hyperparameters used in the SMC algorithm.}
\centering
\begin{tabular}{@{}cccc@{}}
\toprule
\textbf{Definition}                        & \textbf{Symbol}      & \textbf{Num Exp}  & \textbf{Real Exp}  \\ \midrule
Num. samples                          & $N$                  & 50                      & 50 \\
Num. MCMC steps          & $K$                  & 5                       &  5 \\
Variance prop. distr. $\bm{r}_{\rm s}$ & $\sigma^2_{\rm pos}$ & 1          & 1  \\
Variance prop. distr. $\lambda$        & $\sigma^2_{\lambda}$ & 0.2            &  0.01  \\
Variance prop. distr. $\tilde{Q}$      & $\sigma^2_{\tilde{Q}}$       & 25    &  1  \\
Range of values of $\lambda$               & $[\lambda_{\rm m};\lambda_{\rm M}]$     & [0;200] & [0;100] \\
Range of values of $\tilde{Q}$             & $[\tilde{Q}_{\rm m};\tilde{Q}_{\rm M}]$ & [0;$6\cdot10^3$] & [0;$6\cdot10^3$] \\
\bottomrule
\end{tabular}
\label{tab:hyper_SMC}
\end{table}

\begin{algorithm}[H]
\caption{Sequential Monte Carlo with Importance Sampling and perturbation step}
\label{alg:SMC}
\begin{algorithmic}[1]
\State Set initial belief $\tilde{B}^{(0)}$ to a uniform distribution.
\For{$t = 1$ to $T$}
    \State Perform sensors' measurements $\bm{h}^{(t)}$.
    \For{$i = 1$ to $N$}  \Comment{Sample's initialization}
        \State Draw sample from current belief: $\tilde{\bm{\theta}}_i \sim \tilde{B}^{(t-1)}(\bm{\theta})$.
        \State Compute weight $w_i = \mathcal{L}(\bm{h}^{(t)}|\tilde{\bm{\theta}}_i)$.
    \EndFor
    \For{$i = 1$ to $N$}
        \State Normalize weight: $w_i /= \sum\limits_{i=1}^{N} w_i$.
        \State Compute Eff. Sample Size: $N_{{\rm eff}} = 1/\sum\limits_{i=1}^{N} w_i^2$.
    \EndFor
    \If{$N_{\text{eff}} < N_{\text{thr}}$}  \Comment{Resampling step (if needed)}
        \For{$i = 1$ to $N$}
            \State Select $\tilde{\bm{\theta}}_k$ with probability $w_k$.
            \State Put $\tilde{\bm{\Theta}}_i = \tilde{\bm{\theta}}_k$.
        \EndFor
        \For{$i = 1$ to $N$}
            \State Replace $\tilde{\bm{\theta}}_i = \tilde{\bm{\Theta}}_i$.
            \State Set uniform weights: $w_i = 1/N$.
        \EndFor
    \EndIf
    \For{$i = 1$ to $N$}  \Comment{MCMC perturbation step}
        \State Select $l\in\{1,\dots,N\}$ with probability $w_l$.
        \State Set $\bm{\theta}^{(t)}_{i,0} = \tilde{\bm{\theta}}^{(t)}_l$.
        \For{$j = 1$ to $K$}
            \State Sample proposal distr.: $\hat{\bm{\theta}}^{(t)} \sim q(\hat{\bm{\theta}}^{(t)}|\bm{\theta}^{(t)}_{i,j-1})$.
            \State Compute acceptance ratio:
            \State $\alpha=\prod\limits_{k=1}^t \left[\frac{\mathcal{L}(\bm{h}^{(k)}|\hat{\bm{\theta}}^{(k)})}{\mathcal{L}(\bm{h}^{(k)}|\bm{\theta}^{(k)}_{i,j-1})}\right] \frac{q(\bm{\theta}^{(t)}_{i,j-1}|\hat{\bm{\theta}}^{(t)})}{q(\hat{\bm{\theta}}^{(t)}|\bm{\theta}^{(t)}_{i,j-1})}$
            \State Draw random number $u\sim U[0,1]$.
            \If{$u < \min(1, \alpha)$}
                \State Set $\bm{\theta}^{(t)}_{i,j}=\hat{\bm{\theta}}^{(t)}$.
            \Else
                \State Set $\bm{\theta}^{(t)}_{i,j}=\bm{\theta}^{(t)}_{i,j-1}$.
            \EndIf
        \EndFor
        \State Set $\bm{\theta}^{(t)}_i = \bm{\theta}^{(t)}_{i,\tilde{B}}$.
    \EndFor
    \State $\tilde{B}^{(t)}(\bm{\theta}) = \frac{1}{N} \sum\limits_{i=1}^N \delta(\bm{\theta}-\bm{\theta}^{(t)}_i)$. \Comment{Output belief approx.}
\EndFor
\end{algorithmic}
\end{algorithm}

\vfill


%

\end{document}